\documentclass{article}

\usepackage{PRIMEarxiv}

\usepackage[utf8]{inputenc} 
\usepackage[T1]{fontenc}    
\usepackage{hyperref}       
\usepackage{url}            
\usepackage{booktabs}       
\usepackage{amsfonts}       
\usepackage{nicefrac}       
\usepackage{microtype}      
\usepackage{lipsum}
\usepackage{graphicx}
\graphicspath{{media/}}     
\usepackage{algorithm}
\usepackage{algorithmic}
\usepackage{soul}
\usepackage{url}
\usepackage[small]{caption}
\usepackage{graphicx}
\usepackage{amsmath}
\usepackage{amsthm}
\usepackage{booktabs}
\usepackage{algorithm}
\usepackage{algorithmic}
\usepackage[switch]{lineno}
\usepackage{pdfpages}
\usepackage{efbox}
\usepackage{booktabs}

\title{Interactive Melody Generation System for \\ Enhancing the Creativity of Musicians
}

\author{
  So Hirawata, Noriko Otani \\
  Tokyo City University \\
  \texttt{\{hirawata, otani\}@tcu.ac.jp}
}

\begin{document}
\maketitle

\begin{abstract}
This study proposes a system designed to enumerate the process of collaborative composition among humans, using automatic music composition technology. By integrating multiple Recurrent Neural Network (RNN) models, the system provides an experience akin to collaborating with several composers, thereby fostering diverse creativity. 
    Through dynamic adaptation to the user's creative intentions, based on feedback, the system enhances its capability to generate melodies that align with user preferences and creative needs.
    The system's effectiveness was evaluated through experiments with composers of varying backgrounds, revealing its potential to facilitate musical creativity and suggesting avenues for further refinement. The study underscores the importance of interaction between the composer and AI, aiming to make music composition more accessible and personalized. This system represents a step towards integrating AI into the creative process, offering a new tool for composition support and collaborative artistic exploration.
\end{abstract}


\section{Introduction}
In recent years, research on automatic music composition using AI technology has been intensively advanced.
Music composition systems facilitate music production and enable people with no knowledge or composition experience to obtain their own music.
These systems also serves as an educational tool and help beginners learn music theory and composition techniques.
In addition, these have also been used to support composers\cite{otani:gecco}\cite{okabe:jaqp}\cite{hirawata:cmmr}.

Experts in creative fields such must continue to do creative work while making use of their knowledge and skills.
Schön\cite{schon} exlained the importance of ``reflection-in-action'' in creative activities, 
Reflection is triggered by moments of ``surprise'' such as obtaining unexpected results in complex and unpredictable practical situations that experts often face.
Reflection-in-action allows one to recognize that the current approach is not effective and immediately begin to search for new approaches.
The repetition of the cycle of action and reflection brings about changes in artistic expression\cite{yokochi:jcss}.
In addition, encountering a variety of artists promote reflection\cite{okada:jcss}.
Therefore, in creative activities, it is important to obtain ``surprise'' by collaborating with multiple musicians and to continue the transformation through reflection.

Composing music is a task similar to searching for a solution to a problem with a vast solution space, where one has no idea where the solution lies.
Considering that it is a creative activity, it is thought that a richer artistic expression can be realized by working on the solution to this problem in collaboration with others.
In collaborative problem solving, it has been shown that pairs who communicate their opinions to each other tend to get the correct answer more often than pairs who agree with each other\cite{nakano:bps}.
However, it is difficult to give an honest opinion about the work of the collaborators\cite{okabe:jsise}. 
Although some interactive automatic composition systems have been proposed\cite{ando:ieee}\cite{fan:ieee}, which are suitable as collaborators who proceed with the work while expressing open opinions and involving a mutual influence, they are insufficient in terms of obtaining ``surprise'' by collaborating with multiple musicians.

The aim of this study is to support composers in the process of melody generation. A system is proposed that emulates the collaborative composition process with multiple composers, where subsequent melodies are generated based on a melody specified for the beginning.

\section{Related Works}
The work of Otani et al.\cite{otani:gecco} supposed scenarios in which musicians would use an automatic composition system to create music, and proposed the framework for using the composition system in musicians' creative activities and the melody generation method for it.
This system aims to generate musical pieces that adapt to personal sensibilities without imposing any burden on the user.
A new musical piece is generated only by specifying some existing pieces according to the user’s sensibilities, aims, and/or the purpose of the intended composition.
In two cases where the system was used by professional musicians, it was shown that the created music reflected the musicians' sensibilities.

Okabe et al.\cite{okabe:jaqp} demonstrated that this system could enhance the creative resources of musicians and expand their opportunities for reflection.
Since this system operates based on a logic that is completely different from the artists' previous ways of activities, it is thought to have promoted reflection that would have been difficult for a human alone to notice.
In addition, Okabe et al.\cite{okabe:jsise} also showed that collaborative creative activities using this system are effective in the learning of music college students.
In the interview for participants, statements such as ``I am afraid to express my opinion about what the other person is particular about'' and ``I can express my opinion as I like when the computer is the collaborator'' have been obtained.
This shows of the advantages of non-human collaborators.

Furthermore, Hirawata et al.\cite{hirawata:cmmr} developed this system to generate music that includes melodies and chord progressions, and demonstrated its usefulness in a project to create new lullabies based on traditional ones through collaboration between a professional musician and the composition system.
University students who took a class in music creation using this system provided feedback indicating that ``the hurdle of composition'' is significantly lowered as one can select and modify phrases generated by the system, ``the system was perceived as a fellow composer,'' and ``the ability to create a phrase that fits well with a preferred generated phrase was seen as beneficial.''
However, this system is not transformed through interaction with the musician.

The Interactive Composition Aid System by Ando et al.\cite{ando:ieee}, CACIE, which uses interactive evolutionary computation and tree topology for music composition, is recognised for facilitating the creation of complex or atonal pieces of music. However, the explicitness of the composition rules results in a lack of unpredictability.

The interactive automatic composition system proposed by Mo et al.\cite{fan:ieee} uses a transformer-based neural network to generate short segments of music, collecting user feedback to fine-tune the model. This system achieves personalised music composition capabilities that adapt to individual users, but is limited in its ability to modify previous musical information during the dialogue process. Furthermore, the model refined through dialogue is singular and lacks the diversity that could be offered by multiple models.

\section{Melody Generation System}
A simple RNN architecture is used for melody generation, a simple RNN architecture is employed. RNNs remember past data (in this case, previous notes and rhythms) and use this memory to create new musical elements. This feature is particularly effective in capturing the continuity of melodies and rhythmic patterns, and in predicting upcoming notes and phrases.

RNN training is performed using a large dataset of melodies. The network is fed existing melody data and learns to predict the most likely following note for each note. As it does so, the network gradually understands musical patterns, rhythms, and structures, improving its unique melody generation capabilities.

However, when limited melody data is available, fine-tuning pre-trained models can be effective. This approach starts with a general RNN model trained on general music data, and then uses a limited personal dataset to strengthen melody generation in specific styles or genres. This fine-tuning is critical for the model to better adapt to new data and reflect a user's unique melodic style.

\subsection{Melody Generation Method}
Our melody generation system creates melodies in 4/4 time, based on user-entered notes, rests, BPM (beats per minute), and the number of bars. It features multiple melody generation models, each built using a simple RNN (Recurrent Neural Network) architecture. These models can generate continuations of the melody specified by the user. By updating the models' parameters based on user feedback, the system can adapt to the dynamically changing creative intention of the user.

The melody generation process in this system is as follows.

\begin{enumerate}
\item Input determination: The user enters any number of notes and rests, BPM, and the number of bars. Quarter note, eighth note, sixteenth note, quarter rest, eighth rest, and sixteenth rest can be entered.
\item Melody generation: Based on the input musical data, the $N$ melody generation models each generate $M$ melodies.
\item User evaluation: The user rates the $N \times M$ generated melodies on an 11-point scale.
\item Parameter update: The parameters of the models are updated based on the user ratings.
\end{enumerate}

Repeating the above process allows for the generation of melodies that follow to the dynamically changing creative intent of the user. This is similar to a collaborative creation process where the user communicates his impressions to $N$ musicians, who then create new melodies based on his feedback.

\subsection{Parameter Update Method}
The user's ratings for generated melodies are considered performance indicators for the models, 
and the parameters of the melody generation models are updated using this user's ratings.

Particle Swarm Optimization (PSO) is applied to update the parameters.
PSO, an evolutionary computational algorithm inspired by the behavior of biological swarms, 
uses multiple ``particles'' to cooperatively search for the optimal solution.
In our system, a parameter set of a melody generation model is treated as a particle, 
and is changed based on the user ratings of the generated melodies.
To address the slow exploration of PSO, the position update formula is applied $R$ times in a single updating cycle.
This strategy is aimed at accelerating the adaptation of the models to the characteristics of highly rated melodies, thereby reducing the rating burden on human.

Using the rating $v_{tij}$ assigned by the user to the $j$-th melody $melody_{ij}(t)$ generated by the $i$-th model $Model_{i}(t)$ in generation $t$, 
the fitness $fit(Model_{i}(t))$ of $Model_{i}(t)$ is calculated as follows:
\begin{equation}
fit(Model_{i}(t))=\sum_{j}v_{tij}
\end{equation}

After calculating the model's fitness, each model's personal best $pbest_{i}(t)$ and the global best $gbest(t)$ are updated as follows:
\begin{equation}
    \begin{split}
        pbest_{i}(t)&=Model_{i}(t')\\
        &\text{where}\;t'=\mathop{\arg\max}_{t''}fit(Model_{i}(t''))
    \end{split}
\end{equation}
\begin{equation}
    \begin{split}
        gbest(t)&=pbest_{i'}(t)\\
        &\text{where}\;i'=\mathop{\arg\max}_{i''}fit(pbest_{i''}(t))
    \end{split}
\end{equation}
If the user's rating equals the personal best or global best, the personal and global bests are updated to reflect the most recent intention.
The parameters are updated according to the pseudocode in Algorithm\ref{alg:updateParam}, 
which aims to be closer to the global and personal bests compared to standard PSO.
Here, $\vec{g}(t)$ is the global best in generation $t$, 
$\vec{p_i}(t)$ is the personal best of $Model_{i}(t)$, 
$\vec{x_i}(t)$ is the parameter set of $Model_{i}(t)$, 
and $rand[0,1]$ represents a real random number between 0 and 1.

\begin{algorithm}[tb]
\caption{Pseudocode to update the parameters}
\label{alg:updateParam}
    \begin{algorithmic}
    \STATE$\vec{v} = \vec{v_i}(t)$
    \STATE$\vec{x} = \vec{x_i}(t)$
    \FOR{$k=1$ to 50}
    \STATE$\vec{v} = I\vec{v}+A_g\{\vec{g}(t)-\vec{x}\}\times rand[0,1]$
    \STATE$\;\;\;\;\;\;\;\;\;\;\;+A_p\{\vec{p_i}(t)-\vec{x}\}\times rand[0,1]$
    \STATE$\vec{x}=\vec{x}+\vec{v}$
    \ENDFOR
    \STATE$\vec{v_i}(t+1) = \vec{v}$
    \STATE$\vec{x_i}(t+1) = \vec{x}$
    \end{algorithmic}
\end{algorithm}

\subsection{Interface}
The Melody Generation System features a user-friendly interface, central to user interaction and supporting the entire melody generation process.

\begin{figure}[tb]
    \begin{center}
    \includegraphics[scale=0.2]{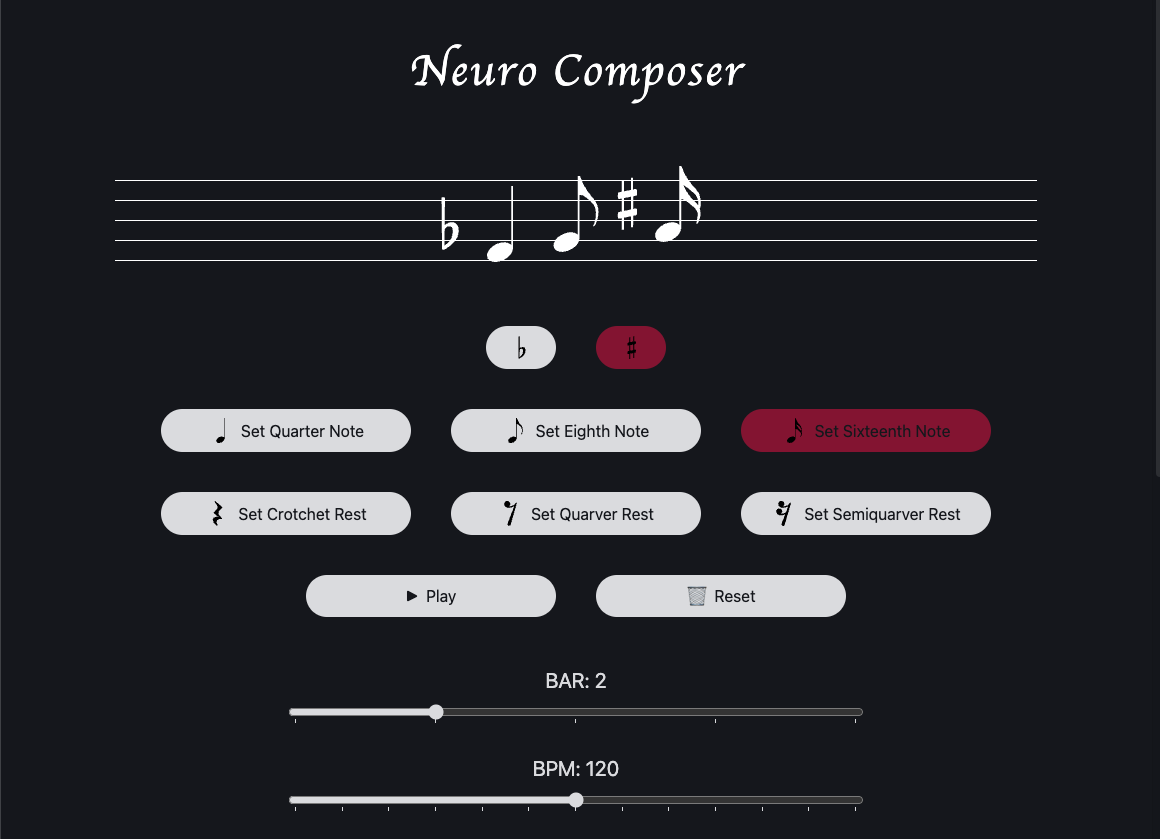}
    \caption{Input configuration screen}
    \label{fig:config}
    \end{center}
\end{figure}

Firstly, users input initial melodies, BPM, and the number of bars using the interface shown in Figure \ref{fig:config}. This initial input forms the foundation of the desired melody. Users can choose any number of notes and rests and specify their preferred BPM and bars, allowing free expression of their musical ideas.

\begin{figure}[tb]
    \begin{center}
    \includegraphics[scale=0.2]{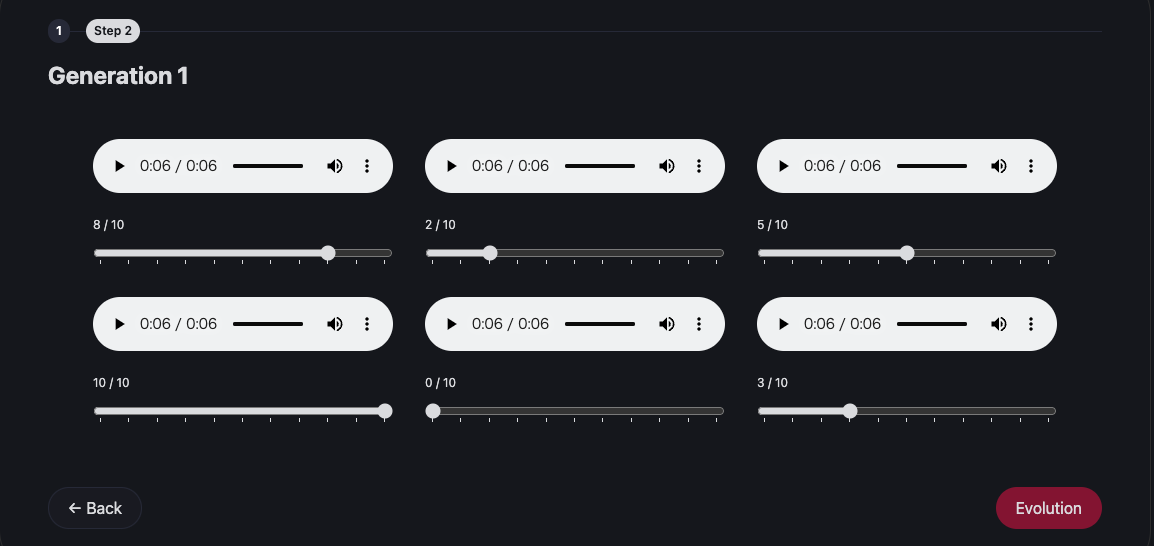}
    \caption{Evaluate melody screen}
    \label{fig:eval}
    \end{center}
\end{figure}

Next, through the interface in Figure \ref{fig:eval}, users listen to melodies generated by the system and evaluate them. Learning from user feedback, the system improves the quality of its melody generation.

These interfaces enable users to intuitively interact, create melodies suited to their preferences, and evolve the system based on their inputs and evaluations. This interactive process allows active user involvement in melody production, adapting flexibly to individual musical preferences and styles.

The Melody Generation System not only capitalizes on current technologies but also has significant potential for future development and expansion. Its design is adaptable to evolving neural network technologies, enhancing melody generation accuracy. By integrating newly developed or improved neural network models, the system can create more refined melodies, enriching the user's musical experience.

\section{Experiment}
An experiment was conducted to investigate the impact of the proposed system on the creative activities of four experts A, B, C, and D with compositional experience.
Participants were instructed to freely compose melodies using the system, with A and B composing independently, while C and D collaborated.
After using the system to their satisfaction, their insights were gathered through semi-structured interviews.

The participants' musical experiences were as follows:
\begin{description}
\item[Participant A] An algorithmic composition expert with a background in classical and jazz music, with choral, guitar, and saxophone skills
\item[Participant B] A professional pop and rock musician with keyboard and drums skills and no experience with automated composition technologies
\item[Participant C] A DJ and composer with a background in dance music, with experience in research on automated composition technologies
\item[Participant D] A composer with a background in various genres including rock and pop, with piano skill and no experience with automated composition technologies
\end{description}

Three melody generation models were used in the evaluation experiment ($M$=3). All models were trained using 45,129 tracks from various genres in the Lakh MIDI Dataset. In addition, two of three were fine-tuned using original jazz and classical music compositions, with 50 melodies each. Two melodies of melody ($N$=2) were generated by each model, and the position update equation was applied 50 times ($R$=50).

The number of bars, the BPM, and the notes and rests as the beginning specified by the participants are shown in Table \ref{tab:config}.
No participants changed configurations during the composition process.
Melodies were generated by Participant A over 7 generations, by Participant B over 6 generations, and by Participant C and Participant D over 5 generations.

\begin{table}[tp]
\centering
\caption{Configuration specified by participants}
\label{tab:config}
\begin{tabular}{c|c|r|l}
    \hline
    & No. of &  & \\
    \multicolumn{1}{c|}{\raisebox{1ex}[0pt]{Participant}} & Bars & \multicolumn{1}{c|}{\raisebox{1ex}[0pt]{BPM}} & \multicolumn{1}{c}{\raisebox{1ex}[0pt]{Notes and rests}}\\
    \hline \hline
    A&4&80&\begin{minipage}{35mm}\vspace*{1mm}\includegraphics[scale=0.3]{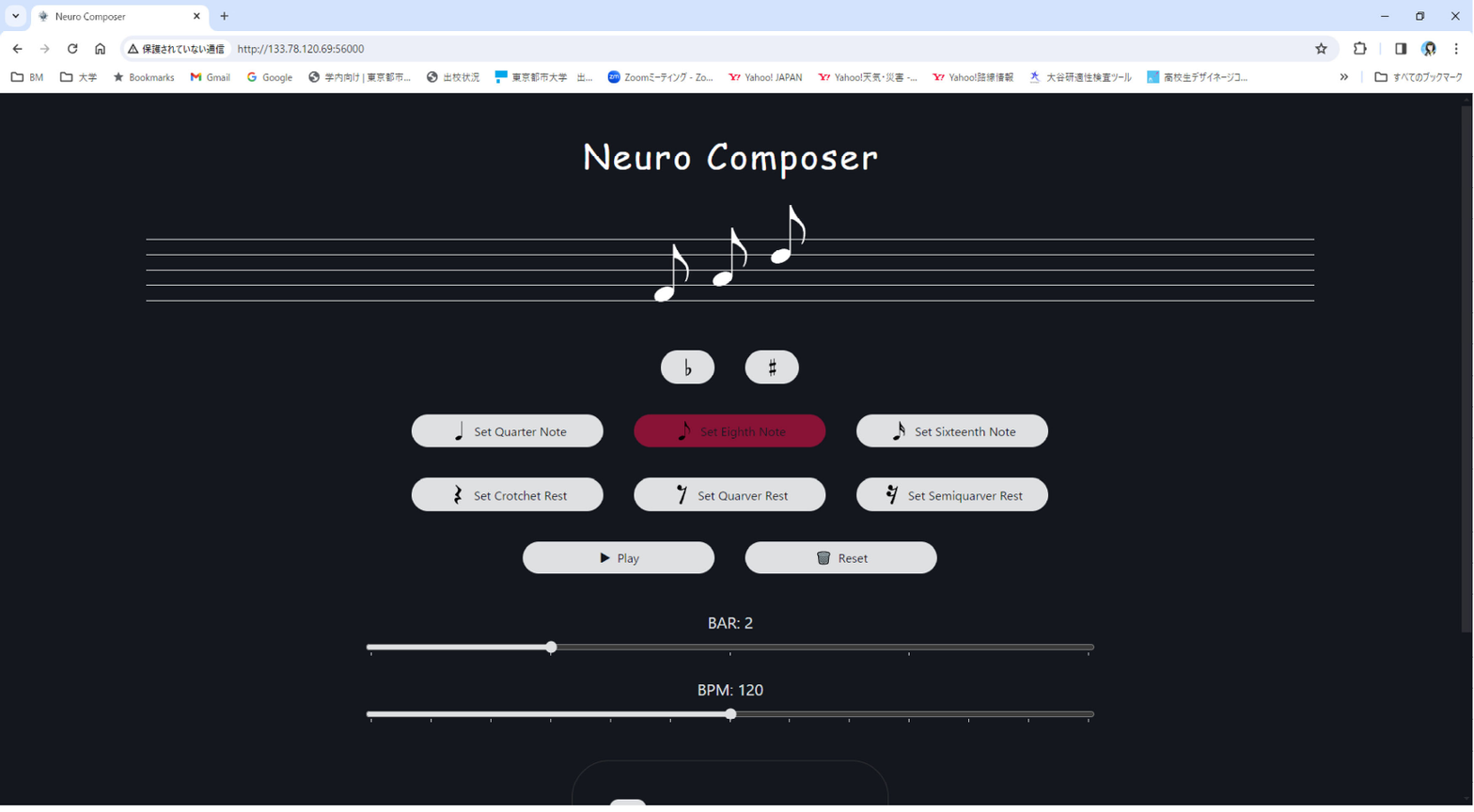}\vspace{1mm}\end{minipage}\\
    \hline
    B&4&120&\begin{minipage}{35mm}\vspace*{1mm}\includegraphics[scale=0.3]{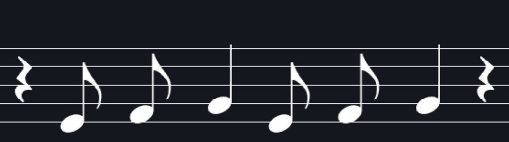}\vspace{1mm}\end{minipage}\\
    \hline
    C \& D&4&100&\begin{minipage}{35mm}\vspace*{1mm}\includegraphics[scale=0.3]{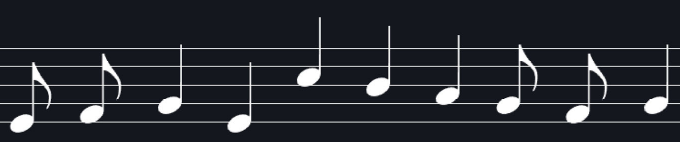}\vspace{1mm}\end{minipage}\\
    \hline
\end{tabular}
\end{table}

\subsection{Analysis of Participants' Statements} 
The participants' statements obtained from the interview were analized using the Steps for Coding and Theorization (SCAT) method, as proposed by Otani, to analyze and theorize qualitative research data\cite{otani:scat}. 
SCAT involves extracting significant codes (keywords or phrases) from the data, which are then combined to form concepts. These concepts are ultimately integrated to construct theories. 
This method proves especially effective in analyzing verbal data, offering deep insights into the data and theoretical understanding.

Table \ref{tab:scat} shows a potion of the result of analysis of Participant B's statements. This illustrates the sequential process of identifying notable phrases in the text, paraphrasing these phrases, describing external concepts that explain these phrases, and considering the context to formulate themes and conceptual frameworks. Theories are then developed by integrating these themes and conceptual frameworks, along with identifying questions and challenges.

Summarizing the theory derived from Participant A's statements, it can be said that the proposed system is adept at balancing traditional musical notation with the exploration of new, creative ideas. This nuanced approach allows composers, especially those in the field of contemporary music, to experiment with novel compositions that incorporate complex rhythms and unique structures, all while remaining grounded in the fundamental aspects of music theory. However, there is a concern that the innovative musical concepts and developments produced by the system may be somewhat challenging for pop music composers and the general public to fully understand.

Participant B's experiment supported the effectiveness of the system in the music production process but suggested that additional features, such as click sounds to aid in rhythm and beat recognition, could enhance its functionality. This suggestion is based on the premise that clicks sounds could make it easier for the general public to recognize the catchiness of music. Furthermore, Participant B was positive about unintended outcomes, especially the emergence of phrases that wouldn't have been thought of otherwise, as positive. According to Schön, reflection is triggered by moments of ``surprise,'' such as unexpected results, which in turn encourages the search for new approaches and solutions. In this way, the system could serve as a catalyst for new creative activities.

The results showed that while the proposed system provides creative opportunities for all participants, it poses comprehension challenges for the general public and pop music composers. Introducing click sounds, as suggested by Participant B, could not only improve the functionality of the system but also enhance the understanding and enjoyment of music by the general public.

The results of the experiments conducted by Participants C and D, in the context of integrating the proposed system into their collaborative music production efforts, suggest that the system may have played a supportive role in enhancing the creative process. 
This inference is drawn from direct quotes from their conversations, which illuminate the potential of the system to nurture a collaborative environment favorable for creative exploration and refinement. For example, Participant C expressed a particular fondness for certain musical elements by saying, ``I really like it when it goes up an octave. That's my favorite,'' reflecting personal preferences that potentially contribute to the collaborative creative process. 
Similarly, Participant D recognized the positive aspects of their collaborative work with remarks such as, ``The beginning was good,'' highlighting instances of consensus and mutual contentment with the musical direction. 
Additional constructive feedback like, ``Yes, yes. That sounds plausible,'' hints at the system's contribution to fostering a shared creative endeavor, where open and constructive dialogue plays a key role in refining and enriching the musical composition.

\begin{table*}[tp]
    \centering
    \caption{Analysis of Participant B's statements}
    \label{tab:scat}
    \begin{tabular}{@{}lp{10cm}@{}}
    \toprule
    \multicolumn{1}{c}{Steps} & \multicolumn{1}{c}{Content} \\ \midrule \midrule \addlinespace 
    Text & Listener: Having actually tried it, how was it? Did you feel it could be used for composing?\\
    & B: Yes, it seems usable. If I were to ask for more, having a click sound accompany the notes would make it more understandable. It's about how the beat is structured; after listening several times, you instantly know which beat melody is playing. \\
    \addlinespace \midrule \addlinespace
    Key Phrases in the Text & Having tried it, usable for composing, if I were to ask for more, having a click sound accompany, how the beat is structured, instantly knowing \\
    \addlinespace \midrule \addlinespace
    Paraphrase of Key Phrases & Felt the possibility of use through direct experience, requested the introduction of a click sound for intuitive feedback, immediate and clear recognition is challenging \\
    \addlinespace \midrule \addlinespace
    Concepts Outside of the Text & Intuitive operability and immediate feedback, visibility of beats and rhythm, enhancement of user experience \\
    \addlinespace \midrule \addlinespace
    Theme and Structural Concept & While the system holds utility in composition, enhancing its functionality requires feedback mechanisms such as click sounds to aid in the recognition of rhythm and beats.\\
    \bottomrule
    \end{tabular}
\end{table*}

\subsection{Analysis of Generated Melodies}
\begin{figure*}[tp]
\fboxsep=0pt
\hspace{3cm}
Model 1 \hspace{2.7cm}
Model 2 \hspace{2.8cm}
Model 3 \hspace{2.9cm}\\
\centering
Gen 1 
\includegraphics[height=15mm]{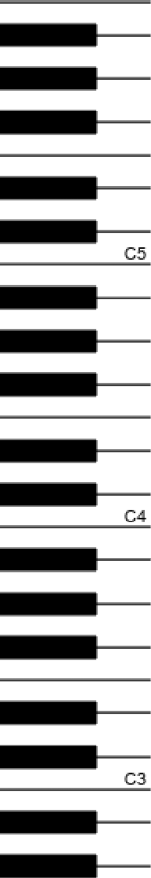}
\fbox{\includegraphics[page=1, height=15mm]{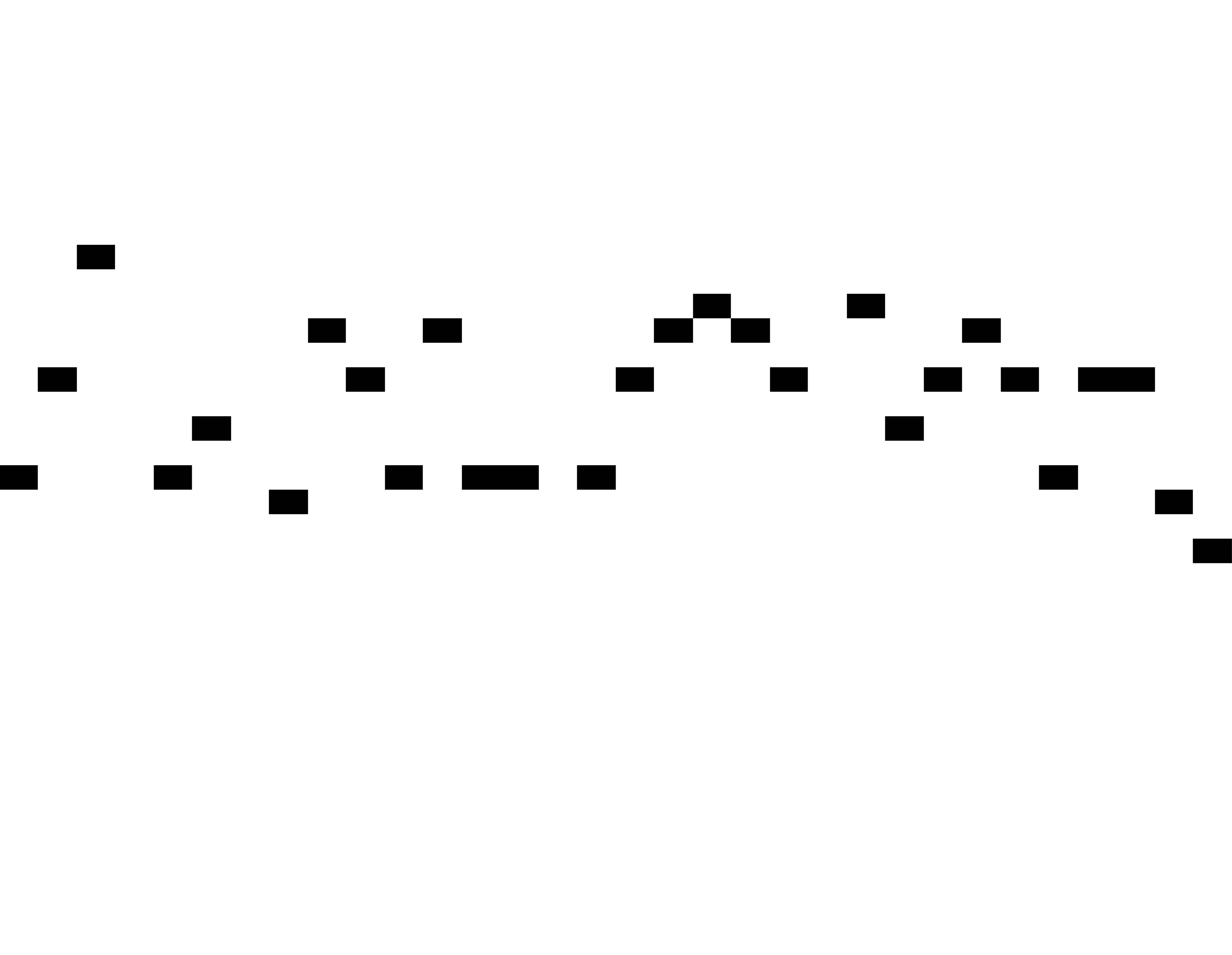}}
\fbox{\includegraphics[page=2, height=15mm]{data/ando/ando1.pdf}}\hspace{1mm}
\fbox{\includegraphics[page=3, height=15mm]{data/ando/ando1.pdf}}
\efbox[linecolor=red]{\includegraphics[page=4, height=15mm]{data/ando/ando1.pdf}}\hspace{1mm}
\fbox{\includegraphics[page=5, height=15mm]{data/ando/ando1.pdf}}
\fbox{\includegraphics[page=6, height=15mm]{data/ando/ando1.pdf}}\\\vspace{0.3mm}
Gen 2 
\includegraphics[height=15mm]{data/piano.pdf}
\fbox{\includegraphics[page=1, height=15mm]{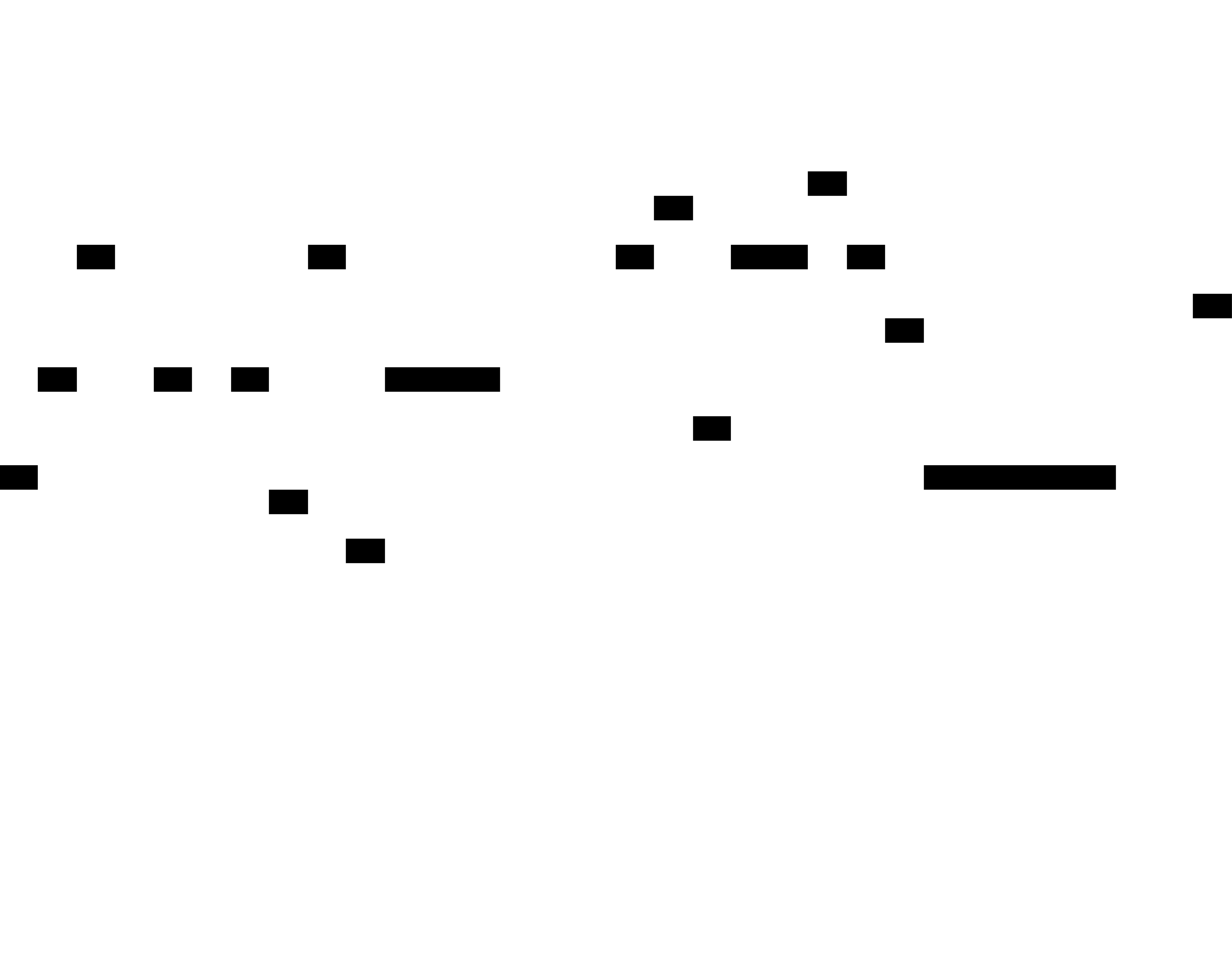}}
\fbox{\includegraphics[page=2, height=15mm]{data/ando/ando2.pdf}}\hspace{1mm}
\fbox{\includegraphics[page=3, height=15mm]{data/ando/ando2.pdf}}
\fbox{\includegraphics[page=4, height=15mm]{data/ando/ando2.pdf}}\hspace{1mm}
\efbox[linecolor=red]{\includegraphics[page=5, height=15mm]{data/ando/ando2.pdf}}
\fbox{\includegraphics[page=6, height=15mm]{data/ando/ando2.pdf}}\\\vspace{0.3mm}
Gen 3 
\includegraphics[height=15mm]{data/piano.pdf}
\fbox{\includegraphics[page=1, height=15mm]{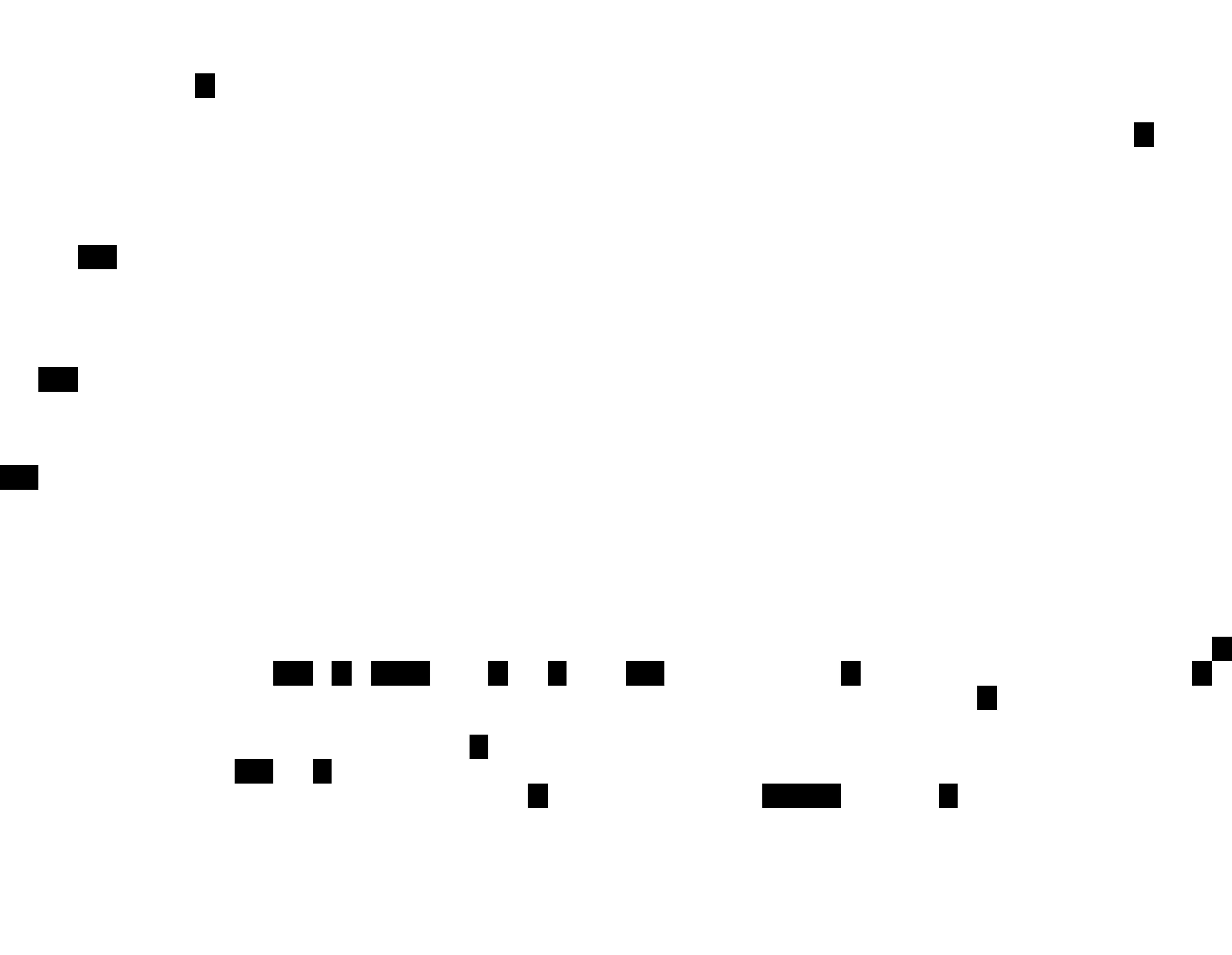}}
\fbox{\includegraphics[page=2, height=15mm]{data/ando/ando3.pdf}}\hspace{1mm}
\efbox[linecolor=red]{\includegraphics[page=3, height=15mm]{data/ando/ando3.pdf}}
\fbox{\includegraphics[page=4, height=15mm]{data/ando/ando3.pdf}}\hspace{1mm}
\fbox{\includegraphics[page=5, height=15mm]{data/ando/ando3.pdf}}
\fbox{\includegraphics[page=6, height=15mm]{data/ando/ando3.pdf}}\\\vspace{0.3mm}
Gen 4 
\includegraphics[height=15mm]{data/piano.pdf}
\fbox{\includegraphics[page=1, height=15mm]{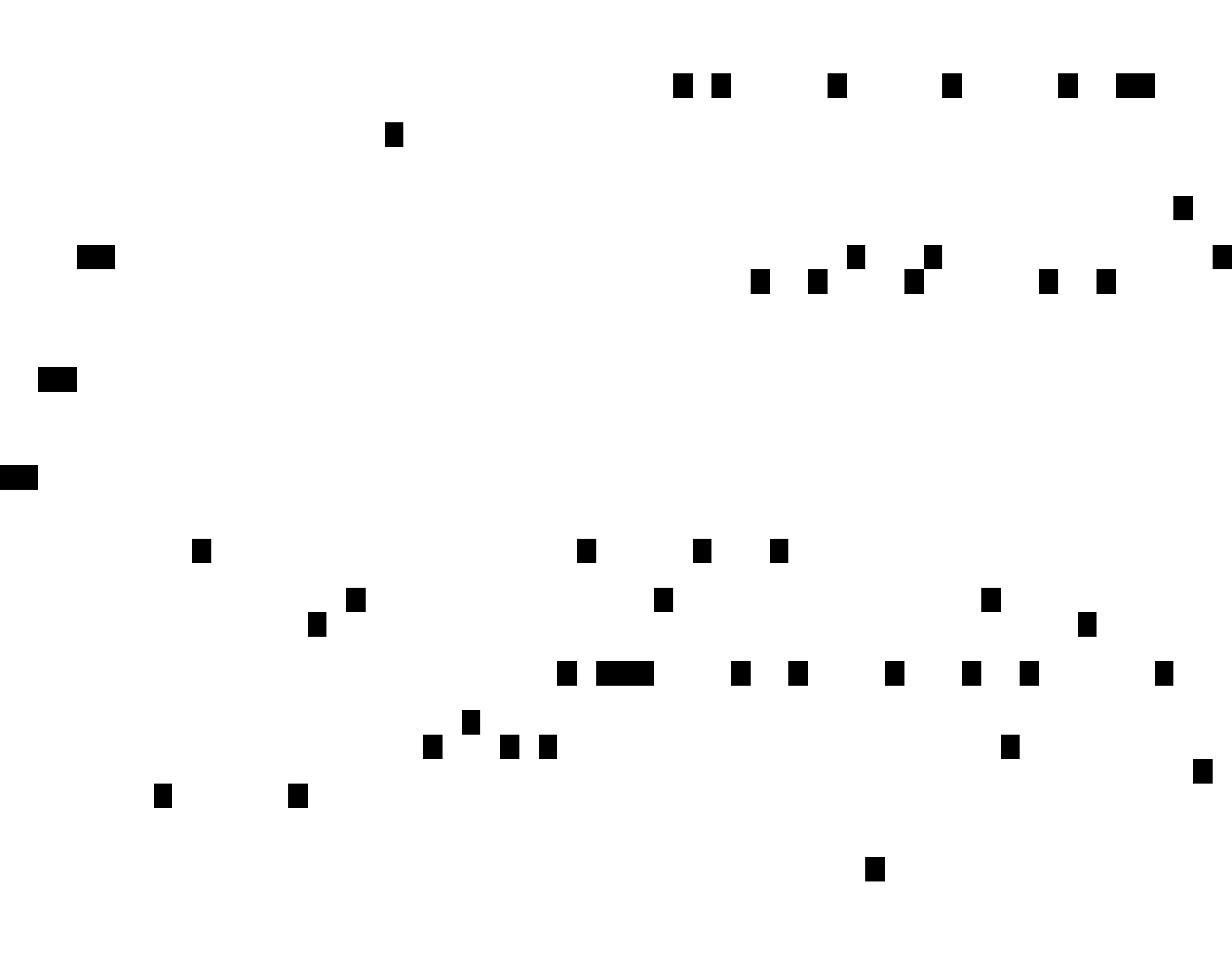}}
\fbox{\includegraphics[page=2, height=15mm]{data/ando/ando4.pdf}}\hspace{1mm}
\fbox{\includegraphics[page=3, height=15mm]{data/ando/ando4.pdf}}
\fbox{\includegraphics[page=4, height=15mm]{data/ando/ando4.pdf}}\hspace{1mm}
\efbox[linecolor=red]{\includegraphics[page=5, height=15mm]{data/ando/ando4.pdf}}
\fbox{\includegraphics[page=6, height=15mm]{data/ando/ando4.pdf}}\\\vspace{0.3mm}
Gen 5 
\includegraphics[height=15mm]{data/piano.pdf}
\efbox[linecolor=red]{\includegraphics[page=1, height=15mm]{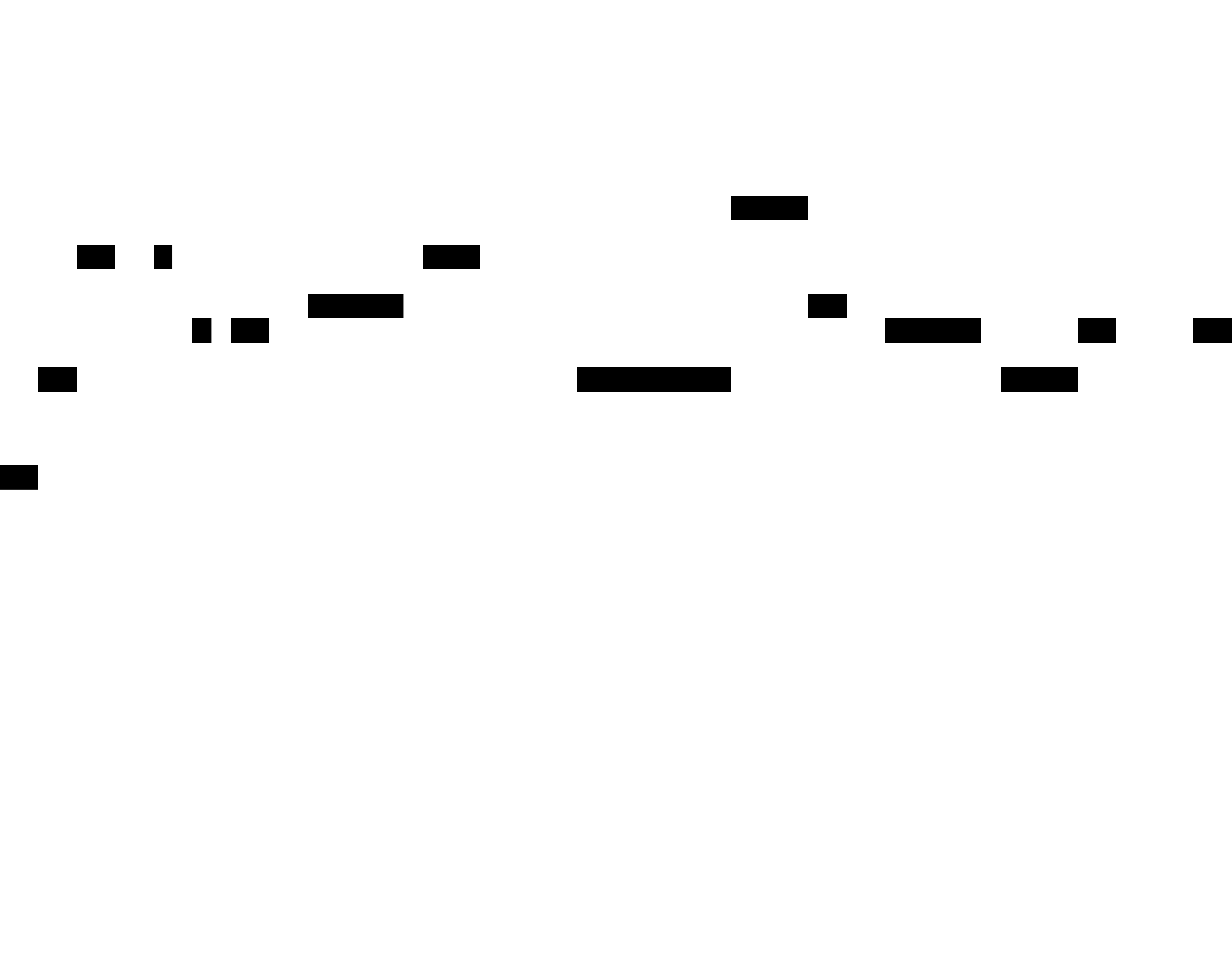}}
\fbox{\includegraphics[page=2, height=15mm]{data/ando/ando5.pdf}}\hspace{1mm}
\efbox[linecolor=red]{\includegraphics[page=3, height=15mm]{data/ando/ando5.pdf}}
\fbox{\includegraphics[page=4, height=15mm]{data/ando/ando5.pdf}}\hspace{1mm}
\fbox{\includegraphics[page=5, height=15mm]{data/ando/ando5.pdf}}
\fbox{\includegraphics[page=6, height=15mm]{data/ando/ando5.pdf}}\\\vspace{0.3mm}
Gen 6 
\includegraphics[height=15mm]{data/piano.pdf}
\fbox{\includegraphics[page=1, height=15mm]{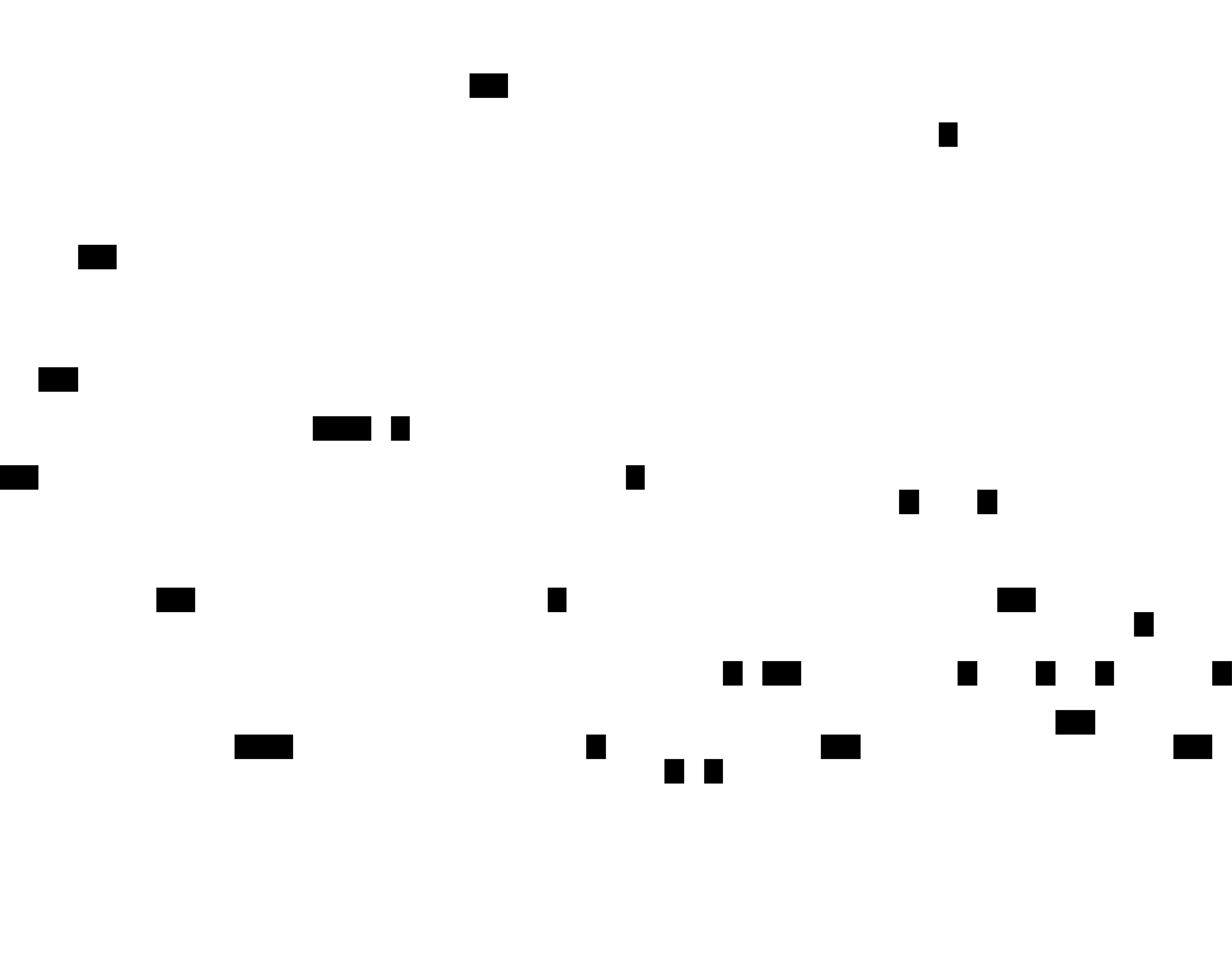}}
\fbox{\includegraphics[page=2, height=15mm]{data/ando/ando6.pdf}}\hspace{1mm}
\fbox{\includegraphics[page=3, height=15mm]{data/ando/ando6.pdf}}
\fbox{\includegraphics[page=4, height=15mm]{data/ando/ando6.pdf}}\hspace{1mm}
\efbox[linecolor=red]{\includegraphics[page=5, height=15mm]{data/ando/ando6.pdf}}
\fbox{\includegraphics[page=6, height=15mm]{data/ando/ando6.pdf}}\\\vspace{0.3mm}
Gen 7 
\includegraphics[height=15mm]{data/piano.pdf}
\fbox{\includegraphics[page=1, height=15mm]{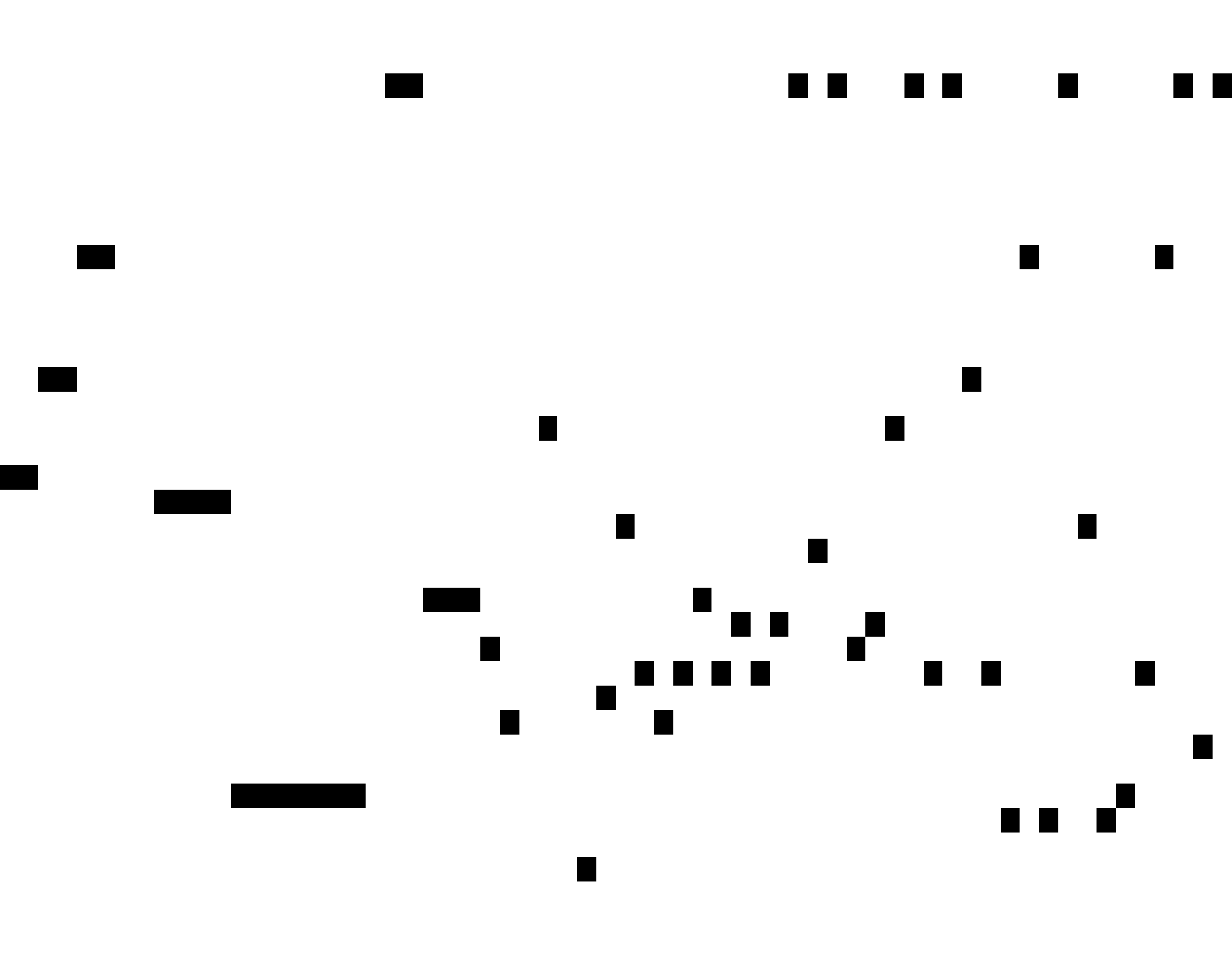}}
\fbox{\includegraphics[page=2, height=15mm]{data/ando/ando7.pdf}}\hspace{1mm}
\fbox{\includegraphics[page=3, height=15mm]{data/ando/ando7.pdf}}
\fbox{\includegraphics[page=4, height=15mm]{data/ando/ando7.pdf}}\hspace{1mm}
\fbox{\includegraphics[page=5, height=15mm]{data/ando/ando7.pdf}}
\fbox{\includegraphics[page=6, height=15mm]{data/ando/ando7.pdf}}\\
(a) Participant A\\\vspace{3mm}
Gen 1 
\includegraphics[height=15mm]{data/piano.pdf}
\fbox{\includegraphics[page=1, height=15mm]{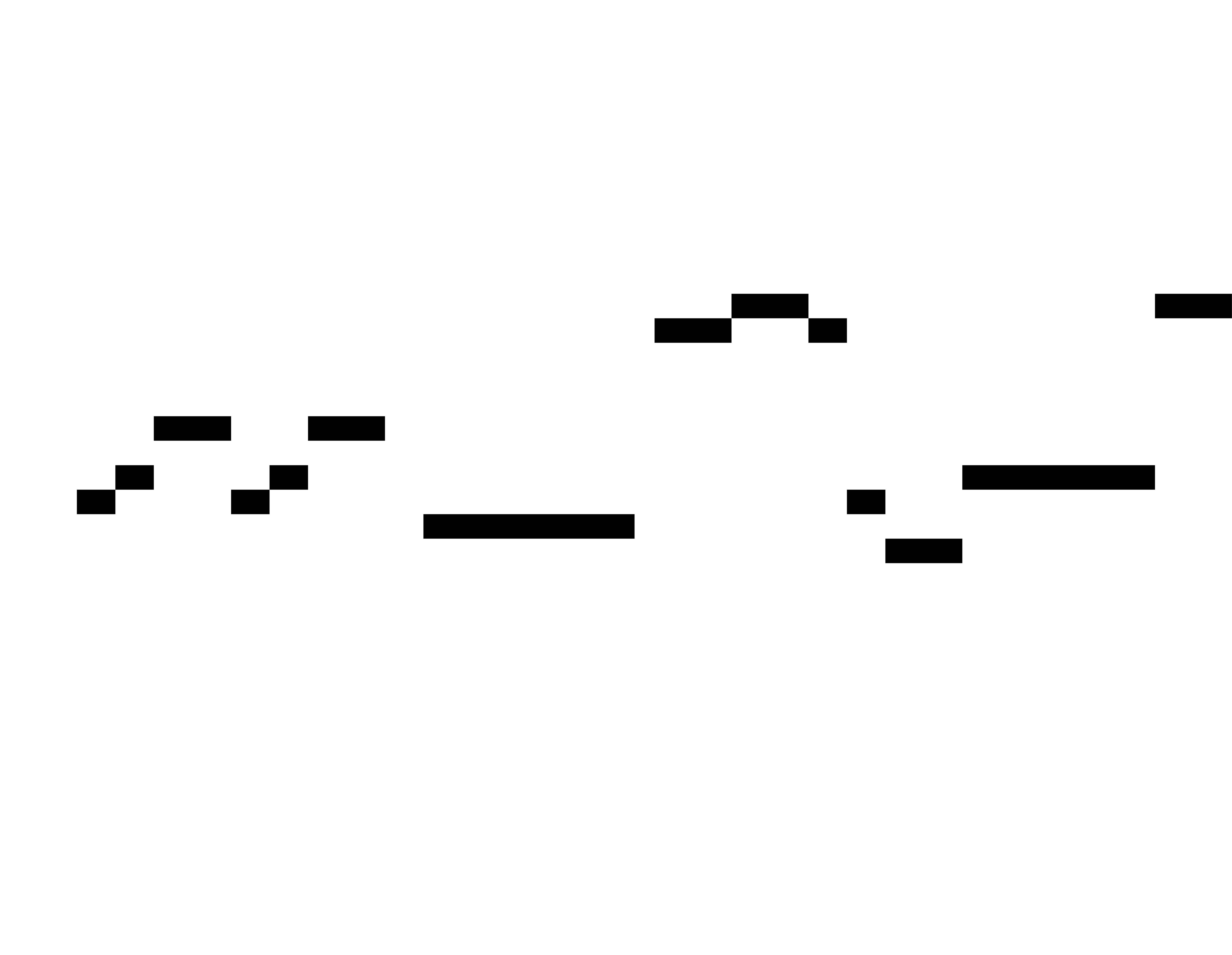}}
\fbox{\includegraphics[page=2, height=15mm]{data/yuuji/yuuji1.pdf}}\hspace{1mm}
\efbox[linecolor=red]{\includegraphics[page=3, height=15mm]{data/yuuji/yuuji1.pdf}}
\fbox{\includegraphics[page=4, height=15mm]{data/yuuji/yuuji1.pdf}}\hspace{1mm}
\fbox{\includegraphics[page=5, height=15mm]{data/yuuji/yuuji1.pdf}}
\fbox{\includegraphics[page=6, height=15mm]{data/yuuji/yuuji1.pdf}}\\\vspace{0.3mm}
Gen 2 
\includegraphics[height=15mm]{data/piano.pdf}
\fbox{\includegraphics[page=1, height=15mm]{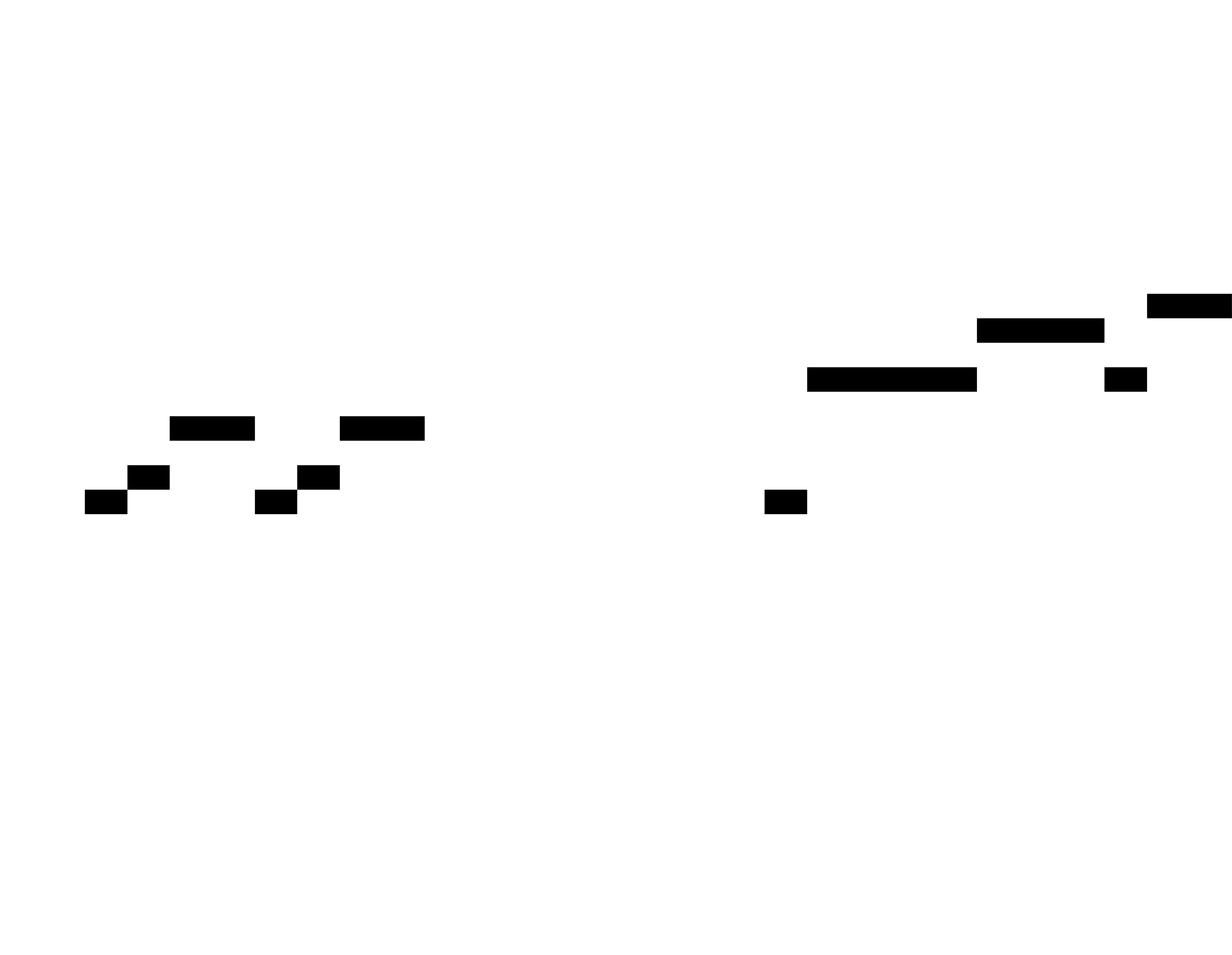}}
\fbox{\includegraphics[page=2, height=15mm]{data/yuuji/yuuji2.pdf}}\hspace{1mm}
\fbox{\includegraphics[page=3, height=15mm]{data/yuuji/yuuji2.pdf}}
\fbox{\includegraphics[page=4, height=15mm]{data/yuuji/yuuji2.pdf}}\hspace{1mm}
\efbox[linecolor=red]{\includegraphics[page=5, height=15mm]{data/yuuji/yuuji2.pdf}}
\fbox{\includegraphics[page=6, height=15mm]{data/yuuji/yuuji2.pdf}}\\\vspace{0.3mm}
Gen 3 
\includegraphics[height=15mm]{data/piano.pdf}
\fbox{\includegraphics[page=1, height=15mm]{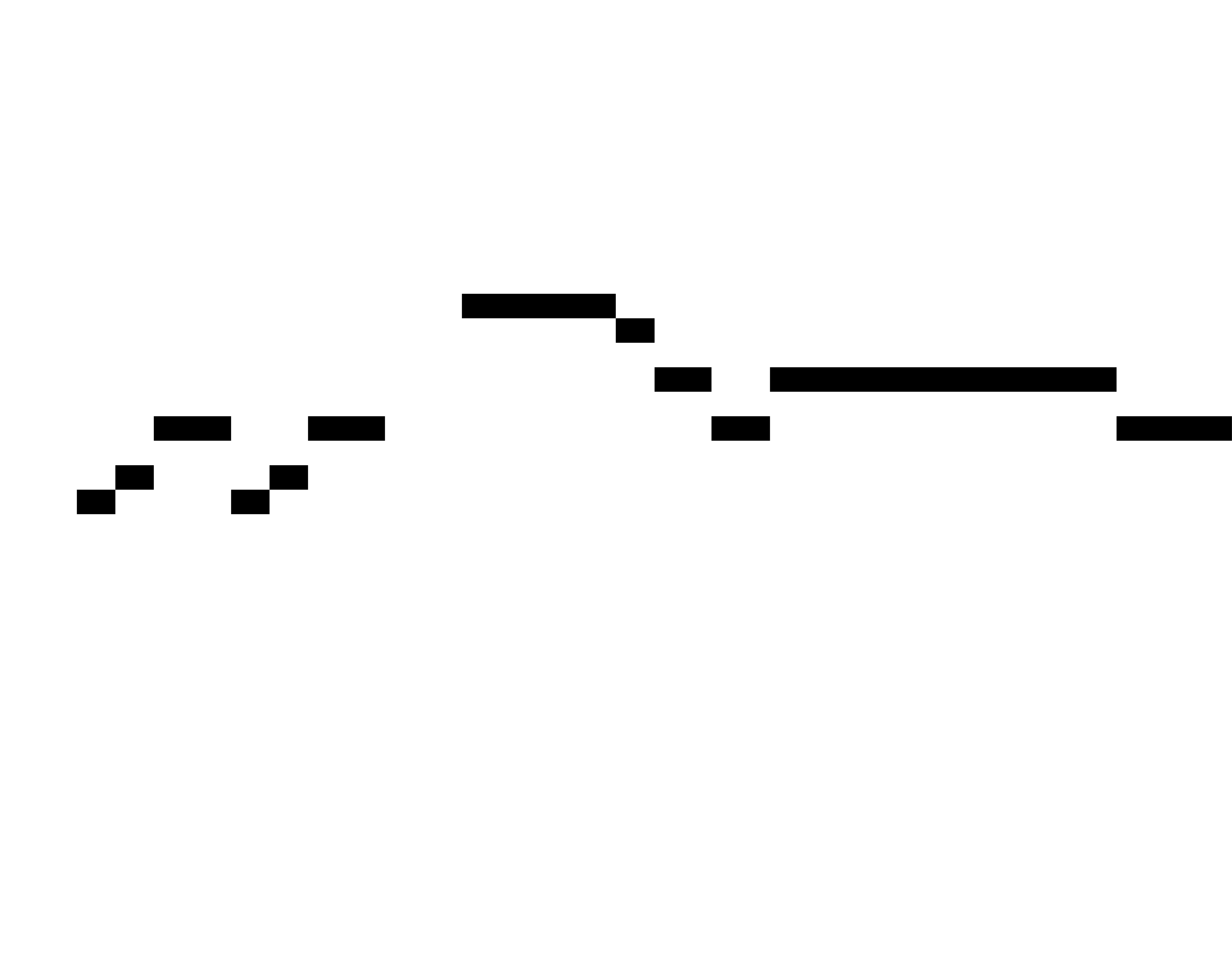}}
\fbox{\includegraphics[page=2, height=15mm]{data/yuuji/yuuji3.pdf}}\hspace{1mm}
\fbox{\includegraphics[page=3, height=15mm]{data/yuuji/yuuji3.pdf}}
\efbox[linecolor=red]{\includegraphics[page=4, height=15mm]{data/yuuji/yuuji3.pdf}}\hspace{1mm}
\fbox{\includegraphics[page=5, height=15mm]{data/yuuji/yuuji3.pdf}}
\fbox{\includegraphics[page=6, height=15mm]{data/yuuji/yuuji3.pdf}}\\\vspace{0.3mm}
Gen 4 
\includegraphics[height=15mm]{data/piano.pdf}
\fbox{\includegraphics[page=1, height=15mm]{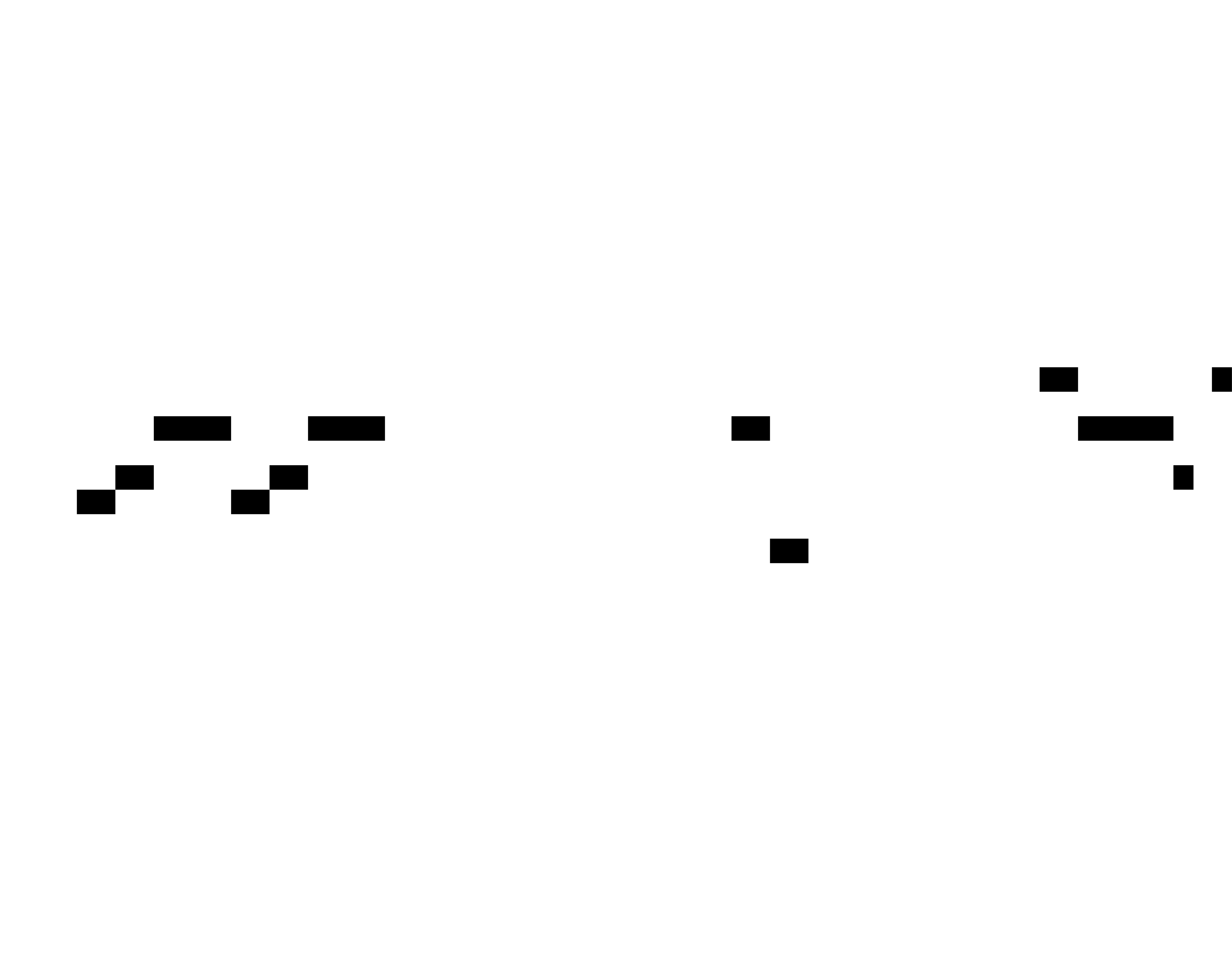}}
\efbox[linecolor=red]{\includegraphics[page=2, height=15mm]{data/yuuji/yuuji4.pdf}}\hspace{1mm}
\fbox{\includegraphics[page=3, height=15mm]{data/yuuji/yuuji4.pdf}}
\fbox{\includegraphics[page=4, height=15mm]{data/yuuji/yuuji4.pdf}}\hspace{1mm}
\efbox[linecolor=red]{\includegraphics[page=5, height=15mm]{data/yuuji/yuuji4.pdf}}
\efbox[linecolor=red]{\includegraphics[page=6, height=15mm]{data/yuuji/yuuji4.pdf}}\\\vspace{0.3mm}
Gen 5 
\includegraphics[height=15mm]{data/piano.pdf}
\fbox{\includegraphics[page=1, height=15mm]{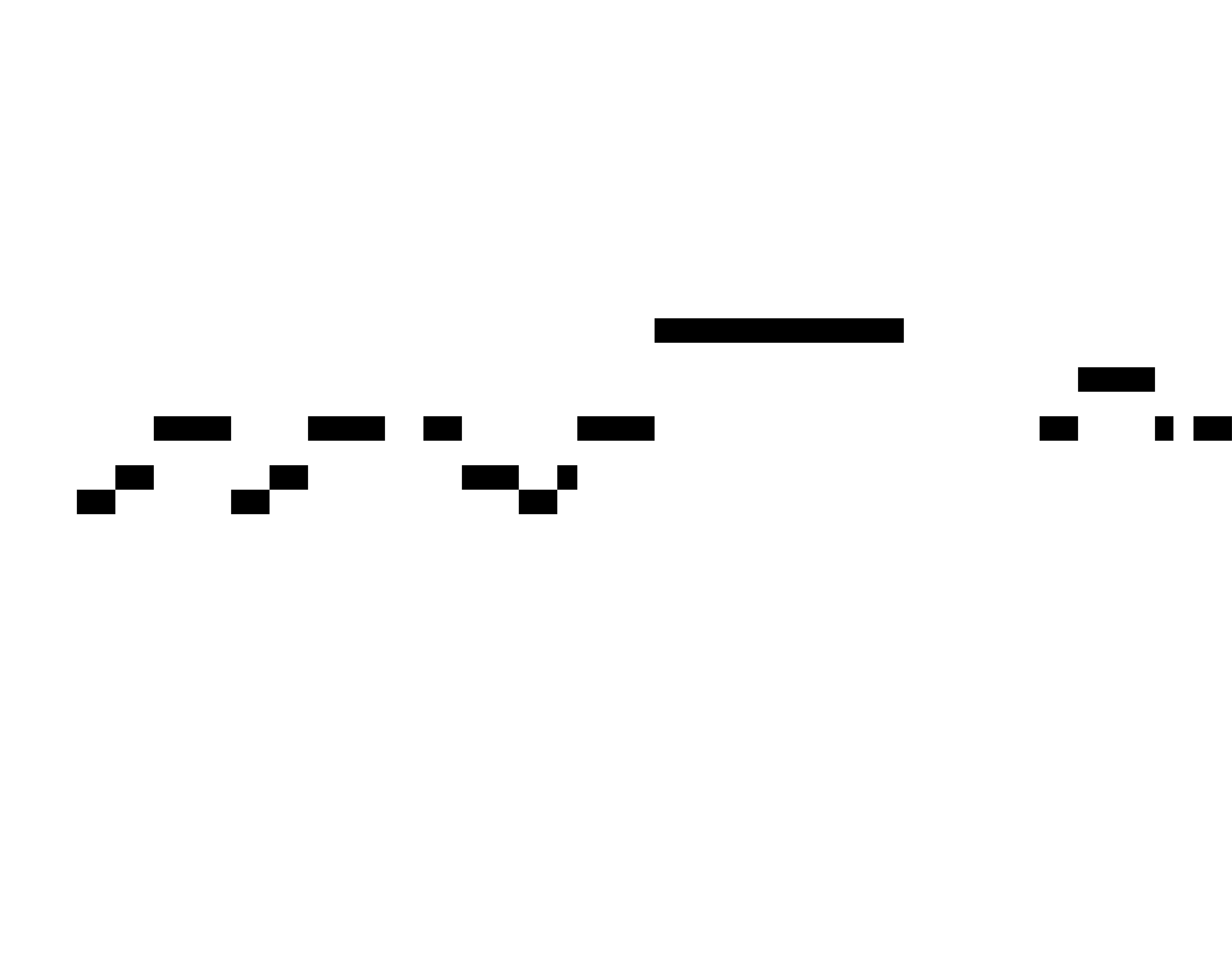}}
\fbox{\includegraphics[page=2, height=15mm]{data/yuuji/yuuji5.pdf}}\hspace{1mm}
\fbox{\includegraphics[page=3, height=15mm]{data/yuuji/yuuji5.pdf}}
\efbox[linecolor=red]{\includegraphics[page=4, height=15mm]{data/yuuji/yuuji5.pdf}}\hspace{1mm}
\fbox{\includegraphics[page=5, height=15mm]{data/yuuji/yuuji5.pdf}}
\fbox{\includegraphics[page=6, height=15mm]{data/yuuji/yuuji5.pdf}}\\\vspace{0.3mm}
Gen 6 
\includegraphics[height=15mm]{data/piano.pdf}
\fbox{\includegraphics[page=1, height=15mm]{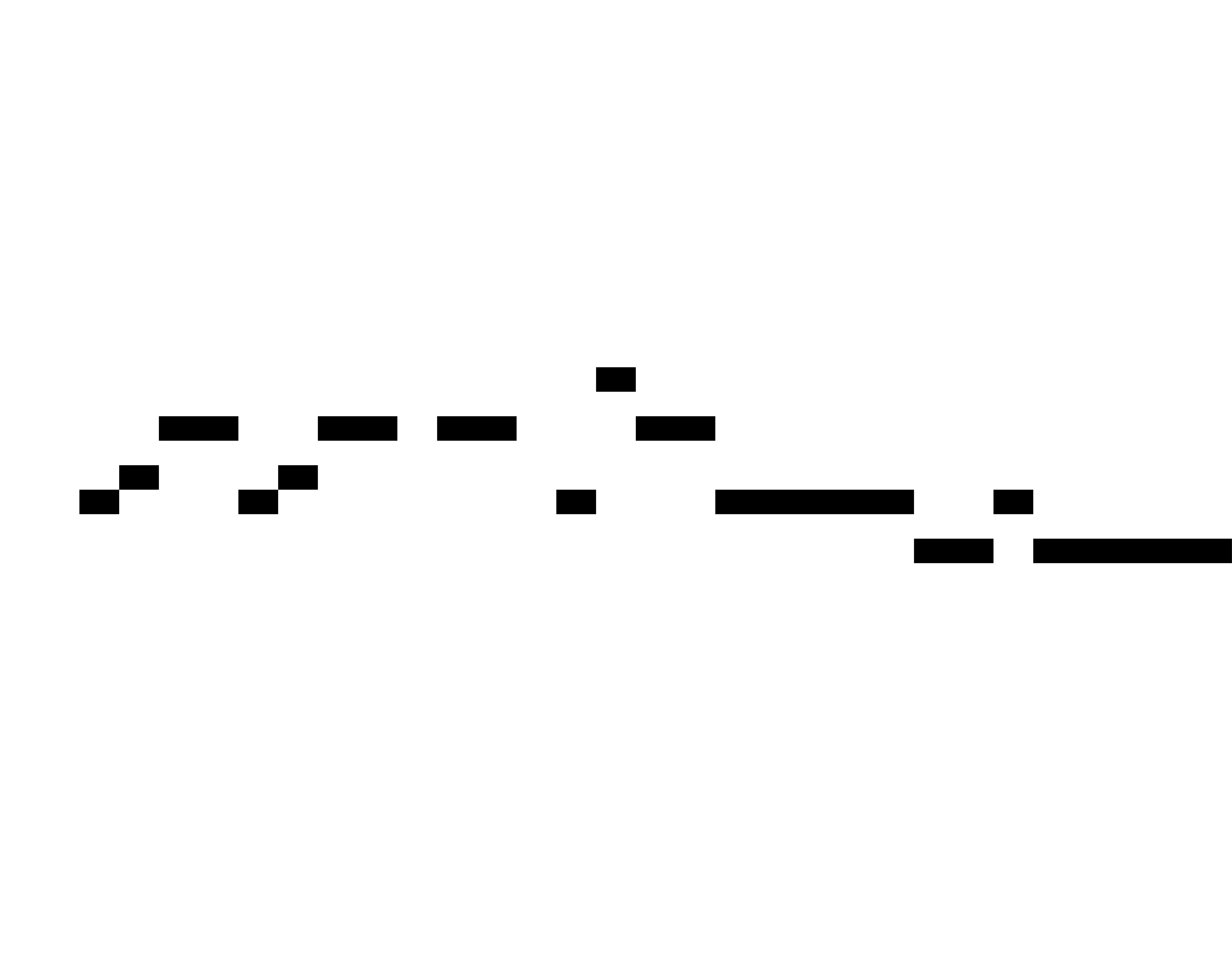}}
\fbox{\includegraphics[page=2, height=15mm]{data/yuuji/yuuji6.pdf}}\hspace{1mm}
\fbox{\includegraphics[page=3, height=15mm]{data/yuuji/yuuji6.pdf}}
\fbox{\includegraphics[page=4, height=15mm]{data/yuuji/yuuji6.pdf}}\hspace{1mm}
\fbox{\includegraphics[page=5, height=15mm]{data/yuuji/yuuji6.pdf}}
\fbox{\includegraphics[page=6, height=15mm]{data/yuuji/yuuji6.pdf}}\\
(b) Participant B
\caption{Piano rolls of generated melodies}
\label{fig:pianoroll}
\end{figure*}

The piano rolls of melody generated by Participant A and Participant B are shown in Figure \ref{fig:pianoroll}.
The two melodies generated by the three models in each generation are on a single line.
The highest rated melody in each generation is shown in the red square.

Figure \ref{fig:pianoroll}(a) shows that the second melody, generated in the first generation with Model 2, was the most highly rated.
This melody contains fewer notes and more notes with larger note values than the other melodies.
This led to an increase in the number of melodies containing many notes with large note values in the second generation, including the second melody from Model 1 and the first melody from Model 2.
However, in the second generation, the first melody generated by Model 3 was the most highly rated.
This melody contains a larger number of notes and many notes with smaller note values than the other melodies.
In response to this, the third generation included more melodies containing many notes with small note values.
The first melody of Model 2, which was the most highly rated in the third generation, and the first melody of Model 3, which was the most highly rated in the fourth generation, had a very narrow range for the fifth note and beyond.
As a result, only melodies with a narrow range were generated in the fifth generation, but because melodies with a relatively large number of notes were highly rated, the sixth generation returned to a situation in which only melodies with a large number of notes were generated.

In the interview, Participant A said ``At first, the two rating criteria were whether or not a melody was generated in accordance with the key from the specified beginning of the melody, and whether or not a rhythm with variations could be generated from three notes of the same note value in the specified beginning. However, after listening to the melodies in the second generation, it became clear that melodies were basically not generated according to the key, and as a specialist in single-note instruments, I began to rate them based on variations in note value and ease of expression. I have changed my rating to that of a person of contemporary music.''
The transition of the melodies in Figure \ref{fig:pianoroll}(a) reflects the statements of Participant A. It can be said that the three collaborative authors in the system, i.e., the melody generation models, follow the dynamic intentions of Participant A.

\begin{figure}[tbp]
\centering
\includegraphics[page=1, scale=1]{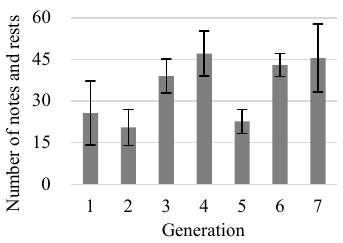}\\
(a) Participant A\\
\includegraphics[page=2, scale=1]{data/noterest.pdf}\\
(b) Participant B
\caption{Average and standard deviation of the number of notes and rests}
\label{fig:noterest}
\end{figure}

Figure \ref{fig:pianoroll}(b) shows that the melody generated by Participant B has less variation in the number of notes and range of pitches than the melody generated by Participant A.
The average and standard deviation of the number of notes and rests in the melody generated in each generation are shown in Figure \ref{fig:noterest}.
If there are many notes and rest with small note values, the number of notes will increase.
In other words, the number of notes and rests can be said to represent the rhythmic tendency.

Participant B generates a melody with the basic premise of singing.
In the interview, he referred to the pitch of the melody by saying, ``It's interesting to hear a melody that surprises me by suddenly jumping to this note after that note.''
Therefore, it can be said that unlike Participant A, who focused on rhythm, Participant B paid more attention to changes in pitch.
The transition of the standard deviation in Figure \ref{fig:noterest} suggests that the melody with the ideal rhythm for Participant B had been generated.

\section{Conclusion}
This study presented a system designed to facilitate melody composition by emulating a collaborative creative process. Utilizing a Recurrent Neural Network (RNN) with feedback-based parameter updates via Particle Swarm Optimization (PSO), the system aims to adapt to and reflect users' creative preferences. Initial feedback from composers indicates the system's potential to assist in musical creativity, suggesting areas for further refinement and user experience enhancement. Future work will focus on improving its melody generation capabilities and integrating user feedback more effectively, with the goal of making the tool more useful and accessible to a wider range of musicians.

\bibliographystyle{unsrt}  
\bibliography{references}

\begin{thebibliography}{10}

\bibitem{otani:gecco}
Noriko Otani, Daisuke Okabe, and Masayuki Numao.
\newblock Generating a melody based on symbiotic evolution for musicians’ creative activities.
\newblock In {\em Proceedings of the Genetic and Evolutionary Computation Conference}, pages 197--204, 2019.

\bibitem{okabe:jaqp}
Daisuke Okabe and Noriko Otani.
\newblock Creative activities and reflection through collaborative composition between professional musician and automatic music composition system.
\newblock {\em Japanese Journal of Qualitative Psychology}, 18:61--75, 2019.

\bibitem{hirawata:cmmr}
So~Hirawata, Noriko Otani, Daisuke Okabe, and Masayuki Numao.
\newblock Creating a new lullaby using an automatic music composition system in collaboration with a musician.
\newblock In {\em Proceedings of the 16th International Symposium on Computer Music Multidisciplinary Research}, pages 524--535, 2023.

\bibitem{schon}
Donald~A. Schön.
\newblock {\em The Reflective Practitioner: How Professionals Think in Action}.
\newblock Basic Books, 1984.

\bibitem{yokochi:jcss}
Sawako Yokochi and Takeshi Okada.
\newblock Creative expertise of contemporary artists.
\newblock {\em Cognitive Studies}, 14(3):437--454, 2007.

\bibitem{okada:jcss}
Takeshi Okada.
\newblock On various aspects of artistic expression: An introduction to the special issue on ``cognitive science of arts''.
\newblock {\em Cognitive Studies}, 20(1):10--18, 2013.

\bibitem{nakano:bps}
Yoshiki Nakano and Keiichi Kodama.
\newblock The effect of collaboration on insightful problem solving in the puzzle of tangram.
\newblock In {\em Book of Abstracts of British Psychology Society - Cognitive Psychology Section Annual Conference 2015}, 2015.

\bibitem{okabe:jsise}
Daisuke Okabe, Noriko Otani, and Yusuke Nagamori.
\newblock A case study of learning environment design for music college students with artificial intelligence.
\newblock {\em Transaction of Japanese Society for Information and Systems in Education}, 37(2):161--166, 2020.

\bibitem{ando:ieee}
Daichi Ando and Hitoshi Iba.
\newblock Interactive composition aid system by means of tree representation of musical phrase.
\newblock In {\em Proceedings of IEEE Congress on Evolutionary Computation}, pages 197--204, 2007.

\bibitem{fan:ieee}
Fan Mo, Xiaoqiang Ji, Huihuan Qian, and Yangsheng Xu.
\newblock A user-customized automatic music composition system.
\newblock In {\em Proceedings of 2022 International Conference on Robotics and Automation}, pages 640--645, 2022.

\bibitem{otani:scat}
Takashi Otani.
\newblock ``scat'' a qualitative data analysis method by four-step coding: easy startable and small scale data-applicable process of theorization [in japanese with english abstract].
\newblock {\em Bulletin of the Graduate School of Education and Human Development}, Educational sciences 54(2):27--44, 2008.

\end{thebibliography}

\end{document}